\documentclass[aps,
showpacs,
twocolumn,
nofootinbib,
nobibnotes,
superscriptaddress,
longbibliography,
amsmath,amssymb,prd]{revtex4-2}

\usepackage[skins,theorems]{tcolorbox}
\usepackage{color}
\usepackage{soul}
\usepackage{centernot}
\usepackage{graphicx}
\graphicspath{{./Figures/}} 
\usepackage{dcolumn}
\usepackage{siunitx}
\usepackage{bm}
\usepackage{bbold}
\usepackage{braket}
\usepackage{esdiff}
\usepackage{hyperref}
\usepackage[mathlines]{lineno}
\usepackage{subfigure}

\begin{document}
\title{Role of indistinguishability and entanglement in Hong-Ou-Mandel interference and finite bandwidth effects of frequency entangled photons}

\author{Roy Barzel}
\affiliation{ZARM, University of Bremen, 28359 Bremen, Germany}
\author{Claus L\"ammerzahl}
\affiliation{ZARM, University of Bremen, 28359 Bremen, Germany}
\affiliation{DLR-Institute for Satellite Geodesy and Inertial Sensing,
c/o University of Bremen, Am Fallturm 2, 28359 Bremen, Germany}
\affiliation{Institute of Physics, Carl von Ossietzky University Oldenburg, 26111 Oldenburg, Germany}


\begin{abstract} 
We investigate the relation between indistinguishability and quantum entanglement in Hong-Ou-Mandel-interference experiments theoretically and relate these quantum mechanical principles to the theorem of entanglement monogamy. Employing Glauber's theory of quantum coherence we compute the detection statistics in HOM-interference of frequency entangled photons, and find a new additional term in the coincidence detection probability, which is related to the spectral indistinguishability of the considered photons that arises from finite bandwidth effects, and therefore is relevant in the limit of low frequency separations or large single-photon bandwidths. Compared to previous work in that context we treat all photonic degrees of freedom (DOF) on equal footing.
\end{abstract} 

\maketitle



\section{\label{sec:introduction} Introduction}
Indistinguishability of physical systems is one of the important defining features of quantum systems without classical analogue. For example, the exchange interaction in metals, which arises as a consequence of the indistinguishability between the electrons in the conduction band, can change the magnetic properties of the system completely \citep{czycholl2007theoretische}. In the case of bosons in particular, indistinguishability leads to intriguing phenomena, such as Bose-Einstein-Condensation or \textit{photon bunching} within interference experiments. The latter has been first witnessed in the pioneering Hong-Ou-Mandel (HOM) experiment \citep{HOM}, and today it is one of the celebrated results of quantum optics that the phenomenon of photon bunching can be used to measure time intervals in sub-pico second regime. This accuracy can be even enhanced in the presence of spectral entanglement \citep{ou1988observation}, giving rise to the yet more counter-intuitive and interesting phenomenon of \textit{photon anti-bunching}, which is very uncharacteristic for photons as these naturally follow bosonic statistics. Needless to be said, quantum entanglement is one of the most important physical phenomena which cannot be explained by means of a classical theory, and moreover its existence has wide consequences in our interpretation of the non-local behavior of matter and the probabalistic character of nature. Since neither photon bunching nor photon anti-bunching can be explained without considering the particle character of photons, and since their occurrence is ultimately related to non-classical features such as indistinguishability and entanglement, which are heralded in a particular way in HOM-interference, these experiments today constitute one of the most important methods to test and study quantum mechanical principles. It is therefore of fundamental interest to precisely understand the role of indistinguishability and entanglement on the detection statistics of HOM-experiments, and the reader is referred to \citep{Bouchard_2020} and references therein for an extensive overview on recent progress and advances in the field, and \citep{2R2R1,2R2R2} for a modern mathematical formulation of the phenomenon in the broader context of multi-particle interference.

In addition the opportunity to provide highly accurate time measurements with HOM-interference entails these experiments also from the practical side a great attraction. Only recently it was demonstrated that HOM-interference can be used to enhance the accuracy of clock synchronization \citep{quan2016demonstration,valencia2004distant}, a feature that was already predicted theoretically  \citep{giovannetti2001quantum}. Therefore, HOM-like schemes are also candidates for space-based implementations, such as the Global Positioning System (GPS), the installation of a global time standard, or in a broader context high precision metrology in general. 

Recently, the possibility to resolve time intervals below $(100\,\mathrm{THz})^{-1}$ (i.e. in the inverse optical regime) with HOM-interference of frequency entangled photons was recognized to bare the potential of conducting fundamental test of physics addressing the interplay between gravity and quantum mechanics \citep{Roura,GIED,BarzelPRD2022}.

In the present paper we employ Glauber's theory of optical coherence \citep{Glauber} and extend previous work on multi-photon quantum interference \citep{pradana2019quantum} to incorporate all photonic degrees of freedom (DOFs) on equal footing. As a result, we obtain a convenient formalism, which on the one hand enables us to compute the detection statistics of multi-photon interference experiments from a general wave function, and on the other hand heralds the role of quantum mechanical principles such as indistinguishability and quantum entanglement in a particular way.

We apply our formalism to compute the detection statistics of HOM interference with frequency entangled photons \citep{Zeillinger} and find a new additional term in the corresponding interference pattern of coincidence detection compared to established results \citep{kaneda2019direct,Zeillinger,ou1988observation}, and provide a physical interpretation for this term.

By contrasting the interference behavior of spectrally indistinguishable frequency entangled photons with the one of spectrally distinguishable frequency detuned photons we draw conclusions on the relation between entanglement and indistinguishability in HOM-interference, and relate these concepts via the theorem of \textit{entanglement monogamy}. 

This work is organized as follows: 
In Section~\ref{sec:HOM} we revise the basics of HOM-interference, present our formalism and apply it to frequency entangled photons. In Section \ref{sec:DISCUSSION} we relate the previously obtained results to former work and provide physical interpretation, and the last Sections \ref{sec:conclusions} and \ref{sec:outlook} are dedicated to conclude our work and provide outlook.


\section{\label{sec:HOM} Hong-Ou-Mandel Interference}
The first two subsections \ref{subsec:TPS} and \ref{subsec:PD} of this chapter are dedicated to develop the theoretical formalism to characterize HOM-interference based on Glauber's theory of coherence. In subsection \ref{subsec:FE} we show one of the main results of the present paper that is the HOM-interference pattern of frequency entangled photons, where we find an additional term related to the finite bandwidth of the employed photons, which was not mentioned in the literature \citep{kaneda2019direct,Zeillinger,ou1988observation} before.

\subsection{Two-photon states}\label{subsec:TPS}
The theory for describing $N$-photon states has been extensively developed \citep{NPhotonState}, and we leave details to the interested reader.
A general two-photon state in the context of HOM-interference is given by \citep{pradana2019quantum}
{\small
\begin{align}
\ket{\psi(\tau_1,\tau_2)}=\mathcal{N}_\psi\int& d \bm{\omega}_1 d \bm{\omega}_2\, \Phi(\bm{\omega}_1,\bm{\omega}_2)e^{i\omega_1\tau_1}e^{i\omega_2\tau_2}
\label{EQ:2phot}
\hat{a}_{\bm{\omega}_1}^\dagger \hat{a}_{\bm{\omega}_2}^\dagger \ket{0},
\end{align}
}
where the bold $\bm{\omega}$ denotes the set of parameters that characterize all degrees of freedom (DOFs) of a single photon, e.g. its frequency, orbital momentum, polarization, spatial mode and so forth. In the case that a discrete DOF is considered (for instance the polarization DOF) the integral in (\ref{EQ:2phot}) has to be evaluated as a sum. We call $\Phi(\bm{\omega}_1,\bm{\omega}_2)$ the two-photon wave function.  The slim $\omega$ denotes the photon frequency, which we treat separately from the other photonic DOFs $\bm{\sigma}$. Thus we can write all photonic DOFs as $\bm{\omega}=\{\omega,\bm{\sigma} \}$. Depending on the context we will alternatively denote the photonic wave function as $\Phi(\bm{\omega}_1,\bm{\omega}_2)=\Phi(\omega_1,\bm{\sigma}_1,\omega_2,\bm{\sigma}_2)=\Phi_{\bm{\sigma}_1\bm{\sigma}_2}({\omega}_1,{\omega}_2)$.

The operators $\hat{a}_{\bm{\omega}},\hat{a}_{\bm{\omega}}^\dagger$ are bosonic annihilation and creation operators respectively, which satisfy the canonical commutator relations $[\hat{a}_{\bm{\omega}},\hat{a}^\dagger_{\bm{\omega}'}]=\delta_{\bm{\omega}\bm{\omega}'}$, while all other commutators vanish. Here $\delta_{\bm{\omega}\bm{\omega}'}$ is the multidimensional delta function of the considered photonic DOFs. For each continuous DOF (for instance the frequency) $\delta_{\bm{\omega}\bm{\omega}'}$ contains a Dirac-delta distribution and for each discrete DOF it contains a Kronecker-symbol as a factor.  For instance if we consider only the photon's polarization $P=H,V$ (horizontal ($H$) and vertical ($V$) polarization) and the photon's frequency $\omega$ (i.e. $\bm{\omega}=\{\omega,P\}$) the delta function reads $\delta_{\bm{\omega}\bm{\omega}'}=\delta(\omega-\omega')\delta_{PP'}$.

\begin{figure}[t]\centering
  \includegraphics[width=1.0\linewidth]{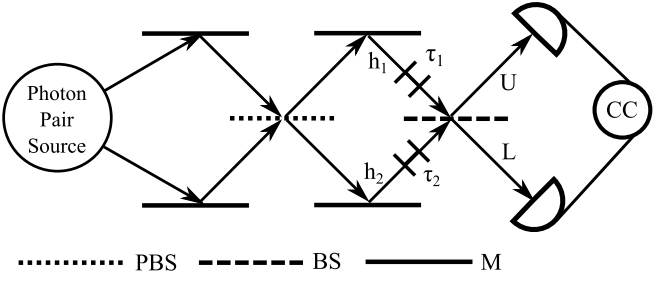}
  \caption{Hong-Ou-Mandel (HOM) experiment with frequency entangled photons. PBS: Polarising beam splitter BS: Beam splitter; M: Mirror; CC: Coincidence count logic.}
\label{fig:HOM}
\end{figure}

Expression \eqref{EQ:2phot} depends on two times $\tau_1$ and $\tau_2$ that describe optical delays that can be applied independently to wave packets in HOM-interferometry \citep{pradana2019quantum}, see Figure~\ref{fig:HOM}. From an experimental point of view, these delays are typically realized via a variation of the optical path of the respective photons. In practice, this can be achieved, for instance, by a variation of the refractive index of the respective transmission path. The normalization constant $\mathcal{N}_\psi$ has to be chosen in
order to fulfill $\braket{\psi|\psi}=1$. However, the procedure that will be described in the following to compute the detection statistics of HOM-interference is independent on the normalization constant as we show further below. For this reason, we omit it for the rest of this work.

\subsection{Photon detection}\label{subsec:PD}
Photon detection is described within Glauber's quantum theory of coherence \citep{Glauber}. The electric
field operators are defined as
\begin{subequations}
\label{EQ:EE}
\begin{align}
\hat{E}^{(+)}_{\bm{\sigma}}(t)=&i\int d{\omega}\,\mathcal{E}_{\omega} e^{-i\omega t}\hat{a}_{\bm{\omega}},\label{EQ:E}
\\
\hat{E}^{(-)}_{\bm{\sigma}}(t)=&(\hat{E}^{(+)}_{\bm{\sigma}}(t))^\dagger,
\end{align}
\end{subequations}
where $\mathcal{E}_\omega=\sqrt{(\hbar\omega/(4\pi\epsilon_0c))}$ is the frequency dependent electric field per photon. The electric field operators are characterized by the parameter set $\bm{\sigma}$.

Analogously we may define creation (annihilation) operators, which create (annihilate) a photon at time $t$, which is characterized by a parameter set $\bm{\sigma}$ as
\begin{subequations}
\label{EQ:a}
\begin{align}
\hat{a}^{}_{\bm{\sigma}}(t)=&\int d\omega e^{-i\omega t}\hat{a}_{\bm{\omega}},\label{EQ:a-}
\\
\hat{a}^{\dagger}_{\bm{\sigma}}(t)=&\int d\omega e^{+i\omega t}\hat{a}_{\bm{\omega}}^\dagger,\label{EQ:a+}
\end{align}
\end{subequations}
The temporal creation (annihilation) operators (\ref{EQ:a}) are basically Fourier transforms w.r.t. the frequency DOF of the spectral creation (annihilation) operators $\hat{a}^{\dagger}_{\omega\bm{\sigma}}$ ($\hat{a}^{}_{\omega\bm{\sigma}}$). Note that the remaining DOFs $\bm{\sigma}$ are not integrated out in (\ref{EQ:a}).

In Glauber's theory of optical coherence \citep{Glauber}, the expectation value of a joint detection of two electric field quanta characterized by $\bm{\sigma}_{1}$ and $\bm{\sigma}_{2}$ at times $t_1$ and $t_2$ in the quantum state (\ref{EQ:2phot}) reads 
\begin{align}
\Gamma_{\bm{\sigma}_1\bm{\sigma}_2}(t_1,t_2)=\braket{\psi|\hat{E}^{(-)}_{\bm{\sigma}_1}(t_1)\hat{E}^{(-)}_{\bm{\sigma}_2}(t_2)\hat{E}^{(+)}_{\bm{\sigma}_2}(t_2)\hat{E}^{(+)}_{\bm{\sigma}_1}(t_1)|\psi},
\end{align}
and the probability $p_{\bm{\sigma}_1\bm{\sigma}_2}(t_1,t_2,\tau_1,\tau_2)$ of a joint detection of two photons characterized by $\bm{\sigma}_1$ and $\bm{\sigma}_2$ at times $t_1$ and $t_2$ is proportional to $\braket{\psi(\tau_1,\tau_2)|\hat{a}^{\dagger}_{\bm{\sigma}_1}(t_1)\hat{a}^{\dagger}_{\bm{\sigma}_2}(t_2)\hat{a}^{}_{\bm{\sigma}_2}(t_2)\hat{a}^{}_{\bm{\sigma}_1}(t_1)|\psi(\tau_1,\tau_2)}\equiv\langle\varphi|\varphi\rangle$, where $\ket{\varphi}:=\hat{a}^{}_{\bm{\sigma}_2}(t_2)\hat{a}^{}_{\bm{\sigma}_1}(t_1)\ket{\psi(\tau_1,\tau_2)}$.
Using (\ref{EQ:2phot}) and (\ref{EQ:a}), together with some algebra, we find that the detection probability $p_{\bm{\sigma}_1\bm{\sigma}_2}(t_1,t_2,\tau_1,\tau_2)\propto\braket{\varphi|\varphi}$ reads
\begin{align}
p_{\bm{\sigma}_1\bm{\sigma}_2}(t_1,t_2,\tau_1,\tau_2)\propto\left|\tilde{\Phi}_{\bm{\sigma}_1\bm{\sigma}_2}(t_1-\tau_1,t_2-\tau_2)\right.+
\nonumber
\\
\left.\tilde{\Phi}_{\bm{\sigma}_2\bm{\sigma}_1}(t_2-\tau_1,t_1-\tau_2)\right|^2,
\end{align}
where we have introduced the Fourier transform of the photonic wave function
\begin{align}
\tilde{\Phi}_{{\bm{\sigma}_1\bm{\sigma}_2}}(T_1,T_2)=\int d\omega_1d\omega_2\ \Phi_{\bm{\sigma}_1\bm{\sigma}_2}(\omega_1,\omega_2)e^{-i\omega_1T_1}e^{-i\omega_2T_2}.
\end{align}
The probability ${P}_{\bm{\sigma}_1\bm{\sigma}_2}(\tau_1,\tau_2)$ to detect two photons irrespectively of the instance of time, when they are detected, is proportional to
\begin{align}
\tilde{P}_{\bm{\sigma}_1\bm{\sigma}_2}(\tau_1,\tau_2):=\int d t_1d t_2\ p_{\bm{\sigma}_1\bm{\sigma}_2}(t_1,t_2,\tau_1,\tau_2).
\end{align} 
We employ the convolution theorem from Fourier analysis and obtain
\begin{align}
\tilde{P}_{\bm{\sigma}_1\bm{\sigma}_2}(\tau_1,\tau_2)=&\int d \omega_1d \omega_2\ \left[|\Phi_{\bm{\sigma}_1\bm{\sigma}_2}(\omega_1,\omega_2)|^2\right.
\nonumber
\\
&\hspace*{20mm} +\left.|\Phi_{\bm{\sigma}_2\bm{\sigma}_1}(\omega_2,\omega_1)|^2\right]
\nonumber
\\
&\hspace*{-20mm} +2\Re\left\{\int d \omega_1d \omega_2 \Phi_{\bm{\sigma}_1\bm{\sigma}_2}(\omega_1,\omega_2)\Phi^*_{\bm{\sigma}_2\bm{\sigma}_1}(\omega_2,\omega_1)\right.
\nonumber
\\
&\times\left. e^{-i(\omega_1-\omega_2)(\tau_1-\tau_2)}\right\},\label{EQ:PX}
\end{align}
which can be further simplified to the compact form
\begin{align}
\tilde{P}_{\bm{\sigma}_1\bm{\sigma}_2}(\tau_1,\tau_2)=\int d \omega_1d \omega_2\ |\mathcal{S}[\Phi^{(\tau_1,\tau_2)}_{\bm{\sigma}_1\bm{\sigma}_2}(\omega_1,\omega_2)]|^2,\label{EQ:P}
\end{align}
where $\Phi^{(\tau_1,\tau_2)}_{\bm{\sigma}_1\bm{\sigma}_2}(\omega_1,\omega_2)=\Phi^{}_{\bm{\sigma}_1\bm{\sigma}_2}(\omega_1,\omega_2)e^{i\omega_1\tau_1}e^{i\omega_2\tau_2}$ is the time dependent photonic wave function, which depends on the optical delays $\tau_{1,2}$, and $\mathcal{S}$ is the symmetrization operator, which acts as $\mathcal{S}[\Phi^{(\tau_1,\tau_2)}_{\bm{\sigma}_1\bm{\sigma}_2}(\omega_1,\omega_2)]=\Phi^{(\tau_1,\tau_2)}_{\bm{\sigma}_1\bm{\sigma}_2}(\omega_1,\omega_2)+\Phi^{(\tau_1,\tau_2)}_{\bm{\sigma}_2\bm{\sigma}_1}(\omega_2,\omega_1)$.
Finally, to obtain the normalized probabilities to detect one photon characterized by $\bm{\sigma}_1$ and another characterized by $\bm{\sigma}_2$ we have to compute
\begin{align}
P_{\bm{\sigma}_1\bm{\sigma}_2}(\tau_1,\tau_2)=\frac{\tilde{P}_{\bm{\sigma}_1\bm{\sigma}_2}(\tau_1,\tau_2)}{\int d\bm{\sigma}_1d\bm{\sigma}_2\, \tilde{P}_{\bm{\sigma}_1\bm{\sigma}_2}(\tau_1,\tau_2)}.\label{EQ:normP}
\end{align}
From the last expression it is seen that the normalization constant $\mathcal{N}_\psi$ from Equation~(\ref{EQ:2phot}) cancels out in the calculation because its square appears in the nominator and denominator of (\ref{EQ:normP}).

Note the symmetries of the probabilities $P_{\bm{\sigma}_1\bm{\sigma}_2}(\tau_1,\tau_2)=P_{\bm{\sigma}_2\bm{\sigma}_1}(\tau_1,\tau_2)=P_{\bm{\sigma}_1\bm{\sigma}_2}(\tau_2,\tau_1)$ and that the probabilities are always a function of the difference $\Delta\tau:=\tau_1-\tau_2$ as can be seen from (\ref{EQ:PX}).

Equation~(\ref{EQ:P}) is a central result of the paper as we can use it to easily compute the joint detection statistics of HOM-interference for any given two-photon wave function $\Phi_{\bm{\sigma}_1\bm{\sigma}_2}({\omega}_1,{\omega}_2)$, and it moreover highlights the interpretation of the absolute square value of the (time dependent) photonic wave function as being the probability for certain detection events. Moreover, Equation~(\ref{EQ:P}) clearly shows that only the symmetrized part of the photonic wave function is of physical relevance, which can be already inferred from Equation (\ref{EQ:2phot}), since it can be written solely in terms of the symmetrized time dependent photonic wave function by use of the canonical commutator relations. 

Apart from that, Equation (\ref{EQ:PX}) emphasizes the role of indistinguishability between the two involved photons in HOM-interference. The first two terms of (\ref{EQ:PX}) do not depend on the delays, in contrast to the last third term, the interference term. This term can be interpreted as the overlap of the joint photonic wave function at the interfering BS with itself under an exchange of the function arguments, i.e. $\bm{\omega}_1 \leftrightarrow \bm{\omega}_2$, i.e. under particle exchange, and thus can serve as a measure of indistinguishability. If this overlap vanishes the particles are considered as entirely distinguishable and HOM-interference does not occur, and we show this on some concrete examples in the later sections.

Moreover, in the absence of entanglement, which is distinguished by the factorization of the two-photon wave function into two single-photon wave functions \citep{ECKERT200288}, (i.e. $\Phi_{\bm{\sigma}_1\bm{\sigma}_2}(\omega_1,\omega_2)=\Phi_{\bm{\sigma}_1}(\omega_1)\Phi_{\bm{\sigma}_2}(\omega_2)$), the interference term of (\ref{EQ:PX}) results in a positive value for $\bm{\sigma}_1=\bm{\sigma}_2$. This increased detection probability for equally behaved ($\bm{\sigma}_1=\bm{\sigma}_2$) photons implies that the factorization of the joint photonic wave function leads to the exclusive occurrence of photon bunching, and thus certifies the occurrence of photon anti-bunching as a sufficient validation for the presence of entanglement, which was already recognized before \citep{HOMTheory}, and also used in experiments \citep{Zeillinger2}. 

Furthermore, we want to add that using the same techniques as in the derivation of (\ref{EQ:P}) we obtain for the single-particle detection statistics which is characterized by the probability $P_{\bm{\sigma}}(\tau_1,\tau_2)=\int dt \braket{\psi(\tau_1,\tau_2)|\hat{a}^{\dagger}_{\bm{\sigma}}(t)\hat{a}^{}_{\bm{\sigma}}(t)|\psi(\tau_1,\tau_2)}$ to detect a single photon in state $\bm{\sigma}$ the intuitive result
\begin{align}
P_{\bm{\sigma}}=\sum_{\bar{\bm{\sigma}}}\frac{1}{2}(P_{\bm{\sigma}\bar{\bm{\sigma}}}+P_{\bar{\bm{\sigma}}\bm{\sigma}})=\sum_{\bar{\bm{\sigma}}}P_{\bm{\sigma}\bar{\bm{\sigma}}}\label{EQ:PSingle}
\end{align}
for all $\tau_1$ and $\tau_2$, which we omitted as function arguments in (\ref{EQ:PSingle}).

Finally, we want to remark that the presented formalism here naturally extends to multi-photon interference experiments, where more than two photons are involved (see for instance \citep{RevModPhys.84.777}). In these cases an analog result to Equation (\ref{EQ:P}) is obtained for the multi-photon detection statistics where one has to symmetrize over the various DOFs of the involved photons. This also includes the special case of single-photon interference experiments like the Mach-Zehnder interferometer. Furthermore, also extending our formalism to mixed states is straightforward. However, we leave the concrete documentation of the extension of our formalism to mixed states for future work. 

\subsection{Frequency Entangled Photons}\label{subsec:FE}
The generation of frequency entangled photons and their subsequent measurement within a HOM-interference experiment is shown in Figure~\ref{fig:HOM} and was first demonstrated in Ref. \citep{ou1988observation}, and recently analyzed in more detail in \citep{Zeillinger}. For this one first generates two polarization entangled and frequency detuned photons which originate from a spontaneously down converting periodically poled $\mathrm{KTiOPO_4}$ crystal (ppKTP-crystal). These photons render (up to normalization) the state 
\begin{align}
\ket{{\psi}_{\mathrm{f.d.}}}
=&\int d {\omega}_1 d {\omega}_2\, {\phi_\mathrm{f.d.}}({\omega}_1,{\omega}_2)e^{i\omega_1\tau_1}e^{i\omega_2\tau_2}\label{EQ:sStateFDt}
\\
&\times\left(\hat{a}_{Hh_1\omega_1}^\dagger \hat{a}_{Hh_2\omega_2}^\dagger + \hat{a}_{Vh_1\omega_1}^\dagger \hat{a}_{Vh_2\omega_2}^\dagger \right)\ket{0},\nonumber
\end{align}
where the spectral wave function in this case reads
\begin{align}
{\phi_\mathrm{f.d.}}(\omega_1,\omega_2)=\delta(\omega_p-\omega_1-\omega_2)\,\mathrm{sinc}\left(\frac{\omega_1-\omega_2-\mu}{\xi}\right).\label{EQ:FE1}
\end{align}
The subscript f.d. means frequency detuned, $\omega_p$ is the pump frequency of the down converting process, $\mu$ is the frequency separation/detuning of the photons (which can be adjusted by tuning the ppKTP-crystal's temperature), and $\xi$ is the single-photon bandwidth. 

The spectral wave function (\ref{EQ:FE1}) is not square integrable due to the occurrence of the delta distribution $\delta(\omega_p-\omega_1-\omega_2)$. This constitutes a problem in the evaluation of Equation (\ref{EQ:PX}) (or rather (\ref{EQ:P})) due to the occurrence of squared delta functions $\delta^2(\omega_p-\omega_1-\omega_2)$ under the integrals to evaluate. One approach to treat this problem stems from scattering theory. There, one uses the identity $\delta(\omega)=\lim_{T\rightarrow \infty}\int_{-T}^T dt\exp(i\omega t)$ to rewrite squares of delta distributions as $\delta^2(\omega)=T\delta(\omega)$, where $T$ is the time over which the various detection events in an experiment are integrated. To obtain particle fluxes instead of particle numbers one has to divide expectation values by $T$. Thus particle fluxes remain finite, also in the limit $T\rightarrow\infty$. The probability of a certain detection event is then recovered through division of the corresponding particle flux of the respective detection event by the sum of particle fluxes of all possible detection events. 

After generation, the two polarization entangled frequency detuned photons interfere on a polarizing beam splitter (PBS), a process that transfers polarization entanglement onto the frequency DOF. The resulting (unnormalized) state after the PBS therefore reads
\begin{align}
\ket{\psi_{\mathrm{f.e.}}}
=&
\int {d {\omega}_1 d {\omega}_2}\, {\phi_\mathrm{f.d.}}({\omega}_1,{\omega}_2)
\label{EQ:freqEnt0}
\\
&\times\left(\hat{a}_{Dh_1\omega_1}^\dagger \hat{a}_{Dh_2\omega_2}^\dagger\right.e^{i\omega_1\tau_1}e^{i\omega_2\tau_2}
\nonumber
\\
&+ e^{i\varphi}\left.\hat{a}_{Dh_1\omega_2}^\dagger \hat{a}_{Dh_2\omega_1}^\dagger e^{i\omega_2\tau_1}e^{i\omega_1\tau_2}\right)\ket{0},
\nonumber
\end{align}
where the subscript f.e. means frequency entangled and the operator $\hat{a}_{Ds\omega}^\dagger=(\hat{a}_{Hs\omega}^\dagger+\hat{a}_{Vs\omega}^\dagger)/\sqrt{2}$ creates a photon in the diagonal polarization state with frequency $\omega$ in the spatial mode $s$. Diagonal polarization of both photons has been achieved by a post-selective measurement in \citep{Zeillinger} (not shown in Figure~\ref{fig:HOM}). The polarization state of the photons is neither manipulated nor filtered or measured after the generation of the frequency entangled photons in the experiment \citep{Zeillinger}. This is why we may discard it from here on, i.e. appart from the frequency we have only the spatial mode as the only left photonic DOF. Thus we have $\bm{\sigma}=s\in\{U,L\}$.

We also included the additional parameter $\varphi$ in the state (\ref{EQ:freqEnt0}). This parameter controls the symmetry of the photonic wave function in the frequency DOF. The value $\varphi=0$ corresponds to a symmetric spectral wave function, while $\varphi=\pi$ corresponds to an anti-symmetric spectral wave function as can be seen further below in Equation~(\ref{EQ:FESpec}).

We can rewrite the state (\ref{EQ:freqEnt0}) as
\begin{align}
\ket{\psi_{\mathrm{f.e.}}}
=&
\int d {\omega}_1 d {\omega}_2 \, \phi^\varphi_\mathrm{f.e.}(\omega_1,\omega_2)e^{i\omega_1\tau_1}e^{i\omega_2\tau_2}
\label{EQ:freqEnt}
\\
&\times \hat{a}_{h_1\omega_1}^\dagger \hat{a}_{h_2\omega_2}^\dagger \ket{0},
\nonumber
\end{align}
where we have defined the frequency entangled spectral wave function
\begin{align}
\phi^\varphi_\mathrm{f.e.}(\omega_1,\omega_2)={\phi_\mathrm{f.d.}}({\omega}_1,{\omega}_2)+e^{i \varphi}{\phi_\mathrm{f.d.}}({\omega}_2,{\omega}_1).\label{EQ:FESpec}
\end{align}

The state \eqref{EQ:freqEnt} then passes through the $50:50$ BS and it transforms into
\begin{align}
\ket{\psi_{\mathrm{f.e.}}}
=&\int {d {\omega}_1 d {\omega}_2}\, \phi^\varphi_\mathrm{f.e.}(\omega_1,\omega_2) e^{i\omega_1\tau_1}e^{i\omega_2\tau_2}
\nonumber
\\
&\times\left(\hat{a}_{U\omega_1}^\dagger \hat{a}_{U\omega_2}^\dagger e^{i\theta} - \hat{a}_{L\omega_1}^\dagger \hat{a}_{L\omega_2}^\dagger e^{-i\theta}\right. \nonumber\\
&\ \ \ + \left.\hat{a}_{U\omega_1}^\dagger \hat{a}_{L\omega_2}^\dagger - \hat{a}_{L\omega_1}^\dagger \hat{a}_{U\omega_2}^\dagger  \right)\ket{0}.
\label{EQ:longState2}
\end{align}

From this we can read off the photonic wave function after the beam splitter $\Phi_{\bm{\sigma}_1\bm{\sigma}_2}({\omega_1},{\omega_2})=\Phi_{s_1s_2}(\omega_1,\omega_2)$, which is characterized by 4 scalar functions in $\omega_1$ and $\omega_2$ (since $s_1$ and $s_2$ respectively can take two values $U,L$). We can arrange the wave function in a $2\times2$-matrix
\begin{align}
{\Phi}_{s_1s_2}(\omega_1,\omega_2)=\phi^\varphi_\mathrm{f.e.}(\omega_1,\omega_2).
\begin{pmatrix}
e^{i\theta} & +1 \\
-1 & -e^{-i\theta}
\end{pmatrix}\label{EQ:redmat2}
\end{align}
The row and column numbering of (\ref{EQ:redmat2}) is $U,V$. 

We now employ (\ref{EQ:redmat2}),(\ref{EQ:P}) and (\ref{EQ:normP}) to compute the joint detection probabilities $\bm{P}_{s_1s_2}(\tau_1,\tau_2)$, and we find
\begin{align}
\bm{P}_{s_1s_2}(\tau_1,\tau_2)=\frac{1}{4}\left[ \bm{R}+d(\tau_1,\tau_2)
\bm{C}\right].\label{EQ:HOMp}
\end{align}
where the matrices $\bm{R}$ and $\bm{C}$ are given in the $(s_1,s_2)$ basis by
\begin{align}
\bm{R}=
\begin{pmatrix}
1 & 1 \\
1 & 1 
\end{pmatrix},
\,\,\,\,\,
\bm{C}=
\begin{pmatrix}
1 & -1 \\
-1 & 1  
\end{pmatrix},\label{EQ:CR}
\end{align}
while the functional dependence on the delays $\tau_1$ and $\tau_2$ appears only in the function $d(\tau_1,\tau_2)$, which we specify immediately.

Note that we have chosen the bold font for $\bm{P}_{s_1s_2}(\tau_1,\tau_2)$ in (\ref{EQ:HOMp}) to emphasize the fact that it has to be read as a matrix. The different entries are equal to the probabilities of the various detection events. The row and column numbering of (\ref{EQ:HOMp}) is the same as in (\ref{EQ:redmat2}). For instance the entry of row two and column two in (\ref{EQ:HOMp}) means that the probability of detecting both photons at detector $L$ (see Figure~\ref{fig:HOM}) is equal to $\bm{P}_{LL}(\tau_1,\tau_2)=1/4(1+d(\tau_1,\tau_2))$. 

The probability $P^{\textrm{c}}(\tau_1,\tau_2):=\bm{P}_{UL}+\bm{P}_{LU}$ for a coincidence measurement (i.e., to detect one photon at one detector $U$ or $L$ and the other photon at the other detector $L$ or $U$) is given by
\begin{align}
P^{\textrm{c}}(\tau_1,\tau_2)=\frac{1}{2}(1-d(\tau_1,\tau_2))\label{EQ:HOM_FE2}  
\end{align}
with
\begin{align}
d(\tau_1,\tau_2)=\frac{R_{\mu\xi}^\varphi(\tau_1,\tau_2)+S_{\mu\xi}(\tau_1,\tau_2)}{1+\cos(\varphi)\mathrm{sinc}\left({2\mu}/{\xi}\right)},\label{EQ:HOM_FE}
\end{align}
where we have introduced the functions
\begin{subequations}
\begin{align}
R_{\mu\xi}^\varphi(\tau_1,\tau_2)=&\cos(\mu(\tau_1-\tau_2)-\varphi)\mathrm{tri}\left(\frac{\xi(\tau_1-\tau_2)}{2}\right),\label{EQ:RF}
\\
S_{\mu\xi}(\tau_1,\tau_2)=&\frac{\sin\left(\frac{2\mu}{\xi}\mathrm{tri}\left(\frac{\xi(\tau_1-\tau_2)}{2}\right)\right)}{\frac{2\mu}{\xi}},\label{EQ:SF}
\end{align}
\end{subequations}
and
\begin{align}
\mathrm{tri}(x)=
\begin{cases}
1-|x| \ \ \ \ \ &\text{  if } |x|\leq 1 \\0 &\text{  if } |x|> 1
\end{cases}
\end{align}
is the \textit{triangular function}.

\section{Discussion}\label{sec:DISCUSSION}
\subsection{Relation to previous experiments}
Our result (\ref{EQ:HOM_FE}) for the HOM-effect extends the scope of application of the one obtained in previous work \citep{Zeillinger}. There, the expression $d(\tau_1,\tau_2)=R_{\mu\xi}^\varphi(\tau_1,\tau_2)$ was obtained, which coincides with our result in the limit $\mu/\xi\gg1$ since $\lim_{\mu_\xi\rightarrow\infty}S_{\mu\xi}(\tau_1,\tau_2)=0$. Let us call $d_0(\tau_1,\tau_2):=  \left. d(\tau_1,\tau_2)\right|_{\mu/\xi\rightarrow\infty}$. Since the previous experiment \citep{Zeillinger} was operated in a parameter regime where one has $\mu/\xi>10$, the discrepancy with our result was not measured. A qualitative difference between our result to the aforementioned result in the literature would have been revealed if the authors had driven the ppKTP-crystal at lower temperatures, but they started their measurement series from the lowest crystal temperature of $T=33.7^\circ\mathrm{C}$, which corresponds to a frequency separation of $\nu=\mu/2\pi=1.7\,\mathrm{THz}$ (see Figure~\ref{fig:Disc}(a)). Nevertheless, also at this frequency separation a slight difference between our result and the one mentioned before can be seen. Our result (\ref{EQ:HOM_FE}) predicts slightly lower interference fringes. This tendency was already seen in the original investigation of frequency entangled photons by Ou and Mandel \citep{ou1988observation}, where this lower fringe visibility was attributed to an imperfect alignment of the measurement apparatus. However, we suggest that this discrepancy might originate, at least in part, due to our correction term $\Delta P^{\textrm{c}}(\tau_1,\tau_2)=d_0(\tau_1,\tau_2)-d(\tau_1,\tau_2)$. In the original derivation \citep{Zeillinger} this lower fringe visibility was explained by imperfect frequency entanglement of the photons.

A clear discrepancy between the approximation formula from the result found in the literature \citep{Zeillinger} and our result for the coincidence probability (\ref{EQ:HOM_FE}) would have been observed if the ppKTP-crystal would have been driven only ten degree Celsius lower at room temperature of about $T\approx20^\circ\,\mathrm{C}$, which would correspond to a frequency separation of $\nu=\mu/2\pi\approx0.265\,\mathrm{THz}$. This would lead to a discrepancy of the coincidence probability to our result of $|\Delta P^{\textrm{c}}(\Delta\tau_{\max})|\approx 0.2$ at a delay of $\Delta\tau_{\max}\approx 1\,\mathrm{ps}$ as seen in Figure~\ref{fig:Disc}(c).

It is also worth noting that in the parameter regime of low frequency separations, the additional term $S_{\mu\xi}(\tau_1,\tau_2)$ predicts qualitatively different physics than those previously obtained via the approximation formula \citep{Zeillinger}, as can be seen in Figure~\ref{fig:Disc}(b) and (c). For a delay of $\Delta\tau_{\max}\approx\pm1.5\,\mathrm{ps}$ our result shows the occurrence of photon bunching (i.e. $P^{\textrm{c}}<0.5$) where the approximation formula $d_0(\tau_1,\tau_2)=R_{\mu\Delta}^\varphi(\tau_1,\tau_2)$ here predicts photon anti-bunching (i.e. $P^{\textrm{c}}>0.5$).

\subsection{Physical interpretation}
The additional term $\Delta P^{\textrm{c}}(\tau_1,\tau_2)\propto S_{\mu\xi}(\tau_1,\tau_2)$ which was not yet documented in the literature \citep{Zeillinger,kaneda2019direct} has a concrete physical meaning as it quantifies the contribution of the spectral indistinguishability related to finite bandwidth effects to the coincidence interference pattern, which becomes more dominant for lower values of $\mu/\xi$ and in particular in the limit $\mu\rightarrow 0$ of vanishingly small frequency separations. 

To see this it is illusive to repeat the calculation of the coincidence detection probability with the spectrum \eqref{EQ:FE1} of frequency detuend photons in place of the spectrum \eqref{EQ:FESpec} of frequency entangled photons. This yields
\begin{align}
P^{\textrm{c}}(\tau_1,\tau_2)=\frac{1}{2}(1-S_{\mu\xi}(\tau_1,\tau_2)),\label{EQ:HOM_FE2X}  
\end{align}
which precisely coincides with the corresponding result of coincidence detection probability of frequency detuned photons found in the literature \citep{Zeillinger2}.
Contrasting this result with the interference pattern 
\eqref{EQ:HOM_FE} of frequency entangled photons shows up the meaning of spectral distinguishability in HOM-interference, and moreover relates this to the presence of entanglement between the spectral DOF and other DOFs of the considered photons, as we explain further below. 

One can regard the spectral wave function (\ref{EQ:FE1}) of frequency detuned photons in the $\omega_1\omega_2$-plane in the limit $\mu\gg\xi$ as a distribution peaked around the point $((\omega_p+\mu)/2,(\omega_p-\mu)/2)$, meaning one photon is emitted at frequency $\omega_1\approx(\omega_p+\mu)/2$ and the other at frequency $\omega_2\approx(\omega_p-\mu)/2$. As can be seen in state (\ref{EQ:sStateFDt}), the photon of frequency $\omega_1$ is emitted in mode $h_1$ and the photon of frequency $\omega_2$ is emitted in mode $h_2$, meaning that here we face a state in which the spatial modes of the photons are entangled with their frequencies. This is why the photons of state \eqref{EQ:sStateFDt} are spectrally distinguishable from each other. The spectral distinguishability increases with increasing values of $\mu/\xi$, i.e. with increasing entanglement between the spectral and spatial DOF of each photon. This suppresses the tendency of frequency detuned photons to bunch or anti-bunch \citep{Zeillinger}, which is reflected in the vanishing of the interference term $S_{\mu\xi}(\tau_1,\tau_2)$ in \eqref{EQ:HOM_FE2X} for increasing values of $\mu/\xi$. 

This emphasizes the role of distinguishability in the context of second quantization, in which no particle can be addressed individually by means of a "label" of the particle. This is the reason why it is often stated that in the second quantization formalism the involved particles are always considered as being indistinguishable. However, the second quantization formalism also carries a notion of distinguishability of particles, which is not encoded in the particle "labels" but in their properties (e.g. DOFs), and the interested reader is referred to \citep{Ghirardi2002,Ghirardi2005} for the original and more detailed discussion on the topic, which was summarized in \citep{RevModPhys.80.517}.

In case of two completely indistinguishable photons of two frequencies $\omega_1$ and $\omega_2$ it is impossible for an experimenter to detect/address a photon of a certain frequency, say $\omega_1$. However, because the property of the photon's frequency (i.e. the photon's frequency DOF) is entangled with its transmission path (i.e. its spatial DOF) in state (\ref{EQ:sStateFDt}), the experimenter can simply place a detector in the transmission path $h_1$ to detect/address the photon of frequency $\omega_1$ with certainty (in the limit $\mu/\xi\rightarrow\infty$).

In contrast to this, the spectrum \eqref{EQ:FESpec} of frequency entangled photons is invariant against exchange of the function arguments $\omega_1$ and  $\omega_2$ for $\varphi=0$. Operationally this means that an experimenter is unable to detect/address a photon of a certain frequency $\omega_1\approx(\omega_p+\mu)/2$ or $\omega_2\approx(\omega_p-\mu)/2$ thorough \textit{any} measurement. This is because (in contrast to the state (\ref{EQ:sStateFDt})) the frequency of the photons is not entangled with their spatial modes (or other DOFs). The only information, which is known with certainty is that at whatever frequency $\omega_1$ or $\omega_2$ one photon of the state (\ref{EQ:freqEnt}) is detected, the other photon is detected at the other frequency $\omega_2$ or $\omega_1$. In other words the two frequencies at which the two photons are detected in state (\ref{EQ:freqEnt}) are known, while the frequency of each individual photon is completely unknown, which is the essence of (spectral) entanglement \citep{Schrodinger}. 

This in turn shows up the relation between distinguishability and entanglement in second quantization. The impossibility to address a certain DOF of single particles (for instance the frequency of a single photon) through \textit{any} measurement corresponds to the indistinguishability of the considered particles w.r.t. this DOF, which is closely related to the presence of entanglement of the considered particles w.r.t. this DOF. However, one should note that the indistinguishability of two particles w.r.t. to a certain property (DOF) is not equivalent but rather a necessary condition for the presence of entanglement in the considered DOF. This means that two photons can be completely indistinguishable in their frequency DOF but nevertheless frequency unentangled (like it is the case with photons from the original HOM-experiment \citep{HOM}). However, the amount of frequency entanglement in a two-photon state is limited by the spectral indistinguishability of the considered photons. 

This can be also seen as a consequence of the theorem of \textit{entanglement-monogamy}, which states that any pair of physical systems which are maximally entangled with each other cannot be entangled with any other physical system and that any entanglement with another, third, physical system comes at the cost of shrinking the amount of entanglement between the first two systems. This is why the interference term $S_{\mu\xi}(\tau_1,\tau_2)$ of the coincidence detection probability (\ref{EQ:HOM_FE2X}) of frequency detuned photons vanishes in the limit $\mu\rightarrow\infty$, because with growing frequency separation $\mu$ the photon frequencies get stronger and stronger entangled with their spatial DOF (i.e. their transmission path), thereby becoming spectrally more and more distinguishable and thus imposing an upper limit on the exploitable frequency entanglement certified by the occurrence of photon anti-bunching \citep{HOMTheory}.

\subsection{Further mathematical considerations}
\begin{widetext}
\begin{figure*}[t]
  \includegraphics[width=0.90\linewidth]{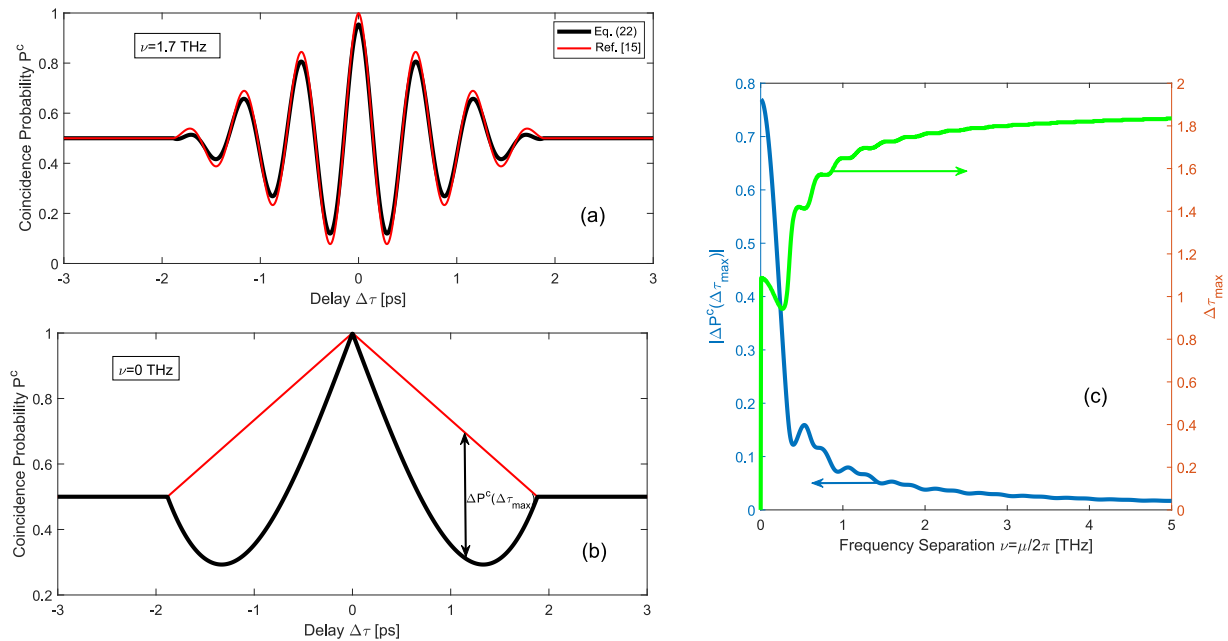}
  \caption{Hong-Ou-Mandel interference experiment with frequency entangled photons (see (\ref{EQ:FESpec}) for joint spectral profile). Coincidence probability for a frequency separation of (a) $\nu=\mu/2\pi=1.7\,\mathrm{THz}$ and (b) $\nu=\mu/2\pi=0\,\mathrm{THz}$. Note that our correction leads to a qualitative different behavior of the coincidence probability. The red curve in (b) predicts photon anti-bunching ($P^{\textrm{c}}>0.5$) for all delays where the black curve predicts photon bunching ($P^{\textrm{c}}<0.5$) for some delays around $\Delta\tau\approx \pm 1.5\,\mathrm{ps}$.  Plot (c) shows with the blue curve the absolute value of the maximum discrepancy $|\Delta P^{\textrm{c}}(\Delta\tau_{\max})|$ in the coincidence probability between our result (Equation~(\ref{EQ:HOM_FE})) and the result, which was obtained in Ref. \citep{Zeillinger} in dependence of the frequency separation $\nu=\mu/2\pi$. The green curve shows the delay $\Delta\tau_{\max}$ at which this discrepancy is maximized. The single-photon bandwidth $\xi=4/\tau_c=1.356\,\mathrm{THz}$ was taken from \citep{Zeillinger}, where the coherence time from the mentioned reference was approximately $\tau_c\approx0.3/0.885\,\mathrm{THz}$. The delay is $\Delta\tau=\tau_1-\tau_2$. All computations were done for $\varphi=\pi$ in our evaluation of (\ref{EQ:HOM_FE}).}
\label{fig:Disc}
\end{figure*}
\end{widetext}

Finally, we want to provide some mathematical arguments why the interference pattern $d_0(\tau_1,\tau_2)$, commonly used in the literature, cannot characterize the coincidence detection statistics of frequency entangled photons in its whole generality. This can be best seen in the limit $\mu\rightarrow 0$. In this limit the spectrum of frequency detunded photons (\ref{EQ:FE1}) is invariant under exchange of the function arguments (i.e. $\phi_{\mathrm{f.d.}}(\omega_1,\omega_2)=\phi_{\mathrm{f.d.}}(\omega_2,\omega_1)$) and thus coincides (apart from a prefactor $(1+e^{i\varphi})$) with the spectrum of frequency entangled photons (\ref{EQ:FESpec}). Thus in the limit $\mu \rightarrow 0$ also the corresponding interference patterns of frequency detuned and frequency entangled photons should coincide, and in particular be independent from the phase $\varphi$ (because it only enters the calculations as a prefactor of the photonic wave function). Note that $\varphi=\pi$ (the case shown in Figure \ref{fig:Disc}(b)) is a special case which we discuss further below separately, i.e. we first consider the case $\varphi \neq \pi$. Indeed our results (\ref{EQ:HOM_FE2}) and (\ref{EQ:HOM_FE2X}) are identical for $\varphi\neq \pi$ and $\mu\rightarrow 0$, and both result in $\lim_{\mu\rightarrow0, \varphi\neq \pi} d(\tau_1,\tau_2)=\mathrm{tri}\left({\xi\Delta\tau}/{2}\right)$ where $d_0(\tau_1,\tau_2)$ does not reproduce the interference pattern of frequency detuned photons (\ref{EQ:HOM_FE2X}) in the limit $\mu\rightarrow 0$ (exept for $\varphi=0$). For $\varphi=\pi/2$ and $\mu\rightarrow 0$ the approximation formula even predicts the absence of interference, i.e. $d_0(\tau_1,\tau_2)=0$. It is interesting to see that for the special case $\varphi=\pi$ our result for the interference terms in the limit $\mu\rightarrow 0$ is $\lim_{\mu\rightarrow0, \varphi= \pi} d(\tau_1,\tau_2)=-\Theta(\xi|\Delta\tau|/2)[(\xi|\Delta\tau|)^3-6\xi|\Delta\tau|+4]/4$, which is shown in Figure \ref{fig:Disc}(b), where $\Theta(x)$ is the Heaviside step function. Despite $\varphi=\pi$ is an interesting mathematical special case in the limit $\mu\rightarrow0$ this special case is physically unstable, since any deviation from the value $\varphi=\pi$ forces the interference term to collapse to the triangular function. Moreover, the special case $\varphi=\pi$ is unphysical in the limit $\mu\rightarrow0$ when one considers frequency entangled photons since in this case the corresponding spectral wave function of frequency entangled photons (\ref{EQ:FESpec}) vanishes. However, we wanted to show this case here (see Figure \ref{fig:Disc}(b)).

\section{\label{sec:outlook} Outlook}
There are several possible extensions to this work.
First of all, we think that our formalism can be easily extended to a much broader class of multi-photon interference experiments \citep{RevModPhys.84.777}, and moreover also to atom-interference experiments \citep{KAUFMAN2018377}. The analysis of many-particle interference with bosons, ferminons or both would be covered by our formalism by a generalized symmetrization operator $\mathcal{S}$ in Equation (\ref{EQ:P}), accounting for the respective parity of the wave function of the quantum system under consideration, and similar results into this direction were already reported in combinatorial approaches \citep{Tichy_2012} and \citep{PhysRevLett.104.220405}. We think that these considerations can be translated into our formalism ans vice versa. However, showing a strict mathematical analogy between these different formalisms remains a subject of future work.

Moreover, a general theoretical framework for the entanglement analysis in the second quantization formalism, which was early recognized \citep{peres2002quantum} to be fundamentally different from the conventional entanglement analysis of distinguishable particles \citep{RevModPhys.81.865}, would be desirable. However, despite tremendous progress \citep{BENATTI20201,RevModPhys.80.517} it is still a subject of ongoing research how to translate key concepts from standard quantum mechanics such as the execution of partial traces and mixed states \citep{lo2016quantum,PhysRevA.99.012341}, various entanglement criteria \citep{ECKERT200288}, or the separability problem in general \citep{PhysRevA.91.042324} to second quantization. 

Furthermore, we want to emphasize that HOM-experiments are ruled by the spectral properties of the employed light sources which are sensitive to one of the trademark predictions from general relativity, namely the redshift on photons propagating through curved spacetime. Moreover, as constituting highly accurate entangled photonic clocks, the study of frequency entangled photons in HOM-interference in a relativistic setting is of great interest. We intend to cover these aspects in a follow-up work. 

\section{\label{sec:conclusions} Conclusions}
Indistinguishability and entanglement are two of the most important genuine quantum mechanical principles without classical analog. Both these principles play a crucial role in HOM-interference experiments which cannot be explained without considering the particle character of single photons which by itself can only be explained within an entirely quantum mechanical theory of electromagnetism. Therefore, HOM-experiments are ideal candidates to study the quantum nature of photons, and moreover the relation between indistinguishability and quantum entanglement.

Using Glauber's theory of optical coherence we developed a formalism to predict the detection statistics of HOM-interference experiments in a systematic way where we treated all photonic DOFs on equal footing under the inclusion of entanglement correlations. This formalism naturally extends also to more complicated experimental setups and multi-particle quantum states. 

We analyzed the role of indistungiushability and entanglement in HOM-interference on the example of two fundamental two-photon sources: Sources which produce spectrally distinguishable frequency detuned photons \citep{Zeillinger2} and sources which produce spectrally indistinguishable frequency entangled photons \citep{Zeillinger}.

The comparison between these sources showed up that the amount of entanglement which can be exploited in HOM-interference through the occurrence of photon anti-bunching is limited by the spectral indistinguishability of the employed photons, which by itself is limited by the amount of entanglement between the spectral DOF of the photons with other DOFs, for instance the spatial one. Therefore, we could relate the relation between indistinguishability and entanglement in HOM-interference to the theorem of entanglement monogamy.

Apart from that, with our formalism we found a new additional term in the interference pattern of frequency entangled photons. Because frequency entangled photons seem to be one of the most promising candidates to test fundamental aspects of physics and moreover they can be used as a resource in quantum technological applications related to high precision metrology it is important to accurately characterize their spectral properties, also in the regime of low frequency separations.

\acknowledgments
We gratefully acknowledge David Edward Bruschi, Andreas Wolfgang Schell and Dennis R\"atzel for fruitful discussions on the topic. R.B. was funded by the Deutsche Forschungsgemeinschaft (DFG, German Research Foundation) under Germany’s Excellence Strategy – EXC-2123 QuantumFrontiers – 390837967. We further gratefully acknowledge support through the TerraQ initiative from the Deutsche Forschungsgemeinschaft (DFG, German Research Foundation) – Project-ID 434617780 – SFB 1464 and the Research Training Group 1620 “Models of Gravity”.

\bibliographystyle{apsrev4-2}
\bibliography{bibtestA}

\appendix

\end{document}